\documentclass[twocolumn]{aastex631}

\begin{document}

\title{Discovery of the richest pulsating ultra-massive white dwarf}

\author{Francisco C. De Gerónimo}
\affiliation{Instituto de Astrof\'{\i}sica de La Plata, CONICET-UNLP, La Plata, Argentina}

\author{Murat Uzundag}
\affiliation{Institute of Astronomy, KU Leuven, Celestijnenlaan 200D, 3001,
    Leuven, Belgium
    }


\author{Alberto Rebassa-Mansergas}
\affiliation{Departament de Física, Universitat Politècnica de Catalunya, c/Esteve Terrades 5, 08860 Castelldefels, Spain}
\affiliation{Institut d'Estudis Espacials de Catalunya, Esteve Terradas, 1, Edifici RDIT, Campus PMT-UPC, 08860 Castelldefels, Barcelona, Spain
}

\author{Alex Brown}
\affiliation{Departament de Física, Universitat Politècnica de Catalunya, c/Esteve Terrades 5, 08860 Castelldefels, Spain}
\author{Mukremin Kilic}
\affiliation{Homer L. Dodge Department of Physics and Astronomy, University of Oklahoma, 440 W. Brooks St., Norman, OK 73019, USA}

\author{Alejandro H. Córsico}
\affiliation{Instituto de Astrof\'{\i}sica de La Plata, CONICET-UNLP, La Plata, Argentina}

\author{Gracyn C. Jewett}
\affiliation{Homer L. Dodge Department of Physics and Astronomy, University of Oklahoma, 440 W. Brooks St., Norman, OK 73019, USA}
\author{Adam G. Moss}
\affiliation{Homer L. Dodge Department of Physics and Astronomy, University of Oklahoma, 440 W. Brooks St., Norman, OK 73019, USA}


\begin{abstract}
The discovery of pulsations in ultra-massive white dwarfs can help to probe their interiors and unveil their core composition and crystallized mass fraction through asteroseismic techniques. To date, the richest pulsating ultra-massive white dwarf known is BPM 37093 with 8 modes detected, for which detailed asteroseismic analysis has been performed in the past.
In this work, we report the discovery of 19 pulsation modes in the ultra-massive white dwarf star WD~J0135+5722, making it the richest pulsating hydrogen-atmosphere ultramassive white dwarf known to date. 
This object exhibits multi-periodic luminosity variations with periods ranging from 137 to 1345 s, typical of pulsating white dwarfs in the ZZ Ceti instability strip, which is centered at $T_{\rm eff} \sim 12\,000$ K. We estimate the stellar mass of WD J0135+5722 by different methods, resulting in $M_{\star} \sim 1.12-1.14 M_{\odot}$ if the star's core is made of oxygen and neon, or $M_{\star} \sim 1.14-1.15 M_{\odot}$ if the star hosts a carbon oxygen  core. Future analysis of the star periods could shed light on the core chemical composition through asteroseismology.

 \end{abstract}



\section{Introduction} 
\label{intro}

ZZ Ceti stars are pulsating hydrogen (H)-rich atmosphere (DAV) white dwarfs (WDs) with effective temperatures in the range $10\,400 {\rm K}\lesssim T_{\rm eff}  \lesssim 13\,000$~K and surface gravities from $\log g\sim 7.5$ to $\sim 9.4$ \citep{2008ARA&A..46..157W, 2008PASP..120.1043F,2010A&ARv..18..471A,2019A&ARv..27....7C,2022PhR...988....1S, 2023MNRAS.522.2181K}. These objects constitute the most common class of pulsating WDs with more than 500 members known to date \citep{2016IBVS.6184....1B,2019A&ARv..27....7C,2020AJ....160..252V,2021ApJ...912..125G, 2022MNRAS.511.1574R, 2024arXiv240707260R}. The pulsations in these objects are multiperiodic, with periods in the range $100 \lesssim \Pi \lesssim 1500$\,s and amplitudes ranging from 0.01 up to 0.3 magnitudes. The variations in  brightness are due to
non-radial g (gravity) modes with low harmonic degree $(\ell \leq 2)$
and generally low to moderate radial order ($1 \lesssim k \lesssim 15$),  excited by the $\kappa-\gamma$ mechanism \citep{1981A&A...102..375D, 1982ApJ...252L..65W} and the convective-driving mechanism
\citep{1991MNRAS.251..673B, 1999ApJ...511..904G}. The existence of the 
red (cool) edge of the ZZ Ceti instability strip can be explained in terms of  excited modes suffering enhanced radiative damping that exceeds 
convective driving, rendering them damped \citep{2018ApJ...863...82L}. 
In many cases, the ZZ Ceti pulsation spectrum exhibits rotational frequency splitting \citep{1975MNRAS.170..405B}, 
which allows constraints on the rotation period \citep[e.g.][]{2017ApJS..232...23H}.

Most WDs are found in the $0.5-1.05$ $M_{\odot}$ range, and those
are expected to have carbon and oxygen (CO) cores \citep{2010A&ARv..18..471A}. Below 0.5 $M_{\odot}$, low-mass WDs
are expected to harbor He-cores \citep{2004ApJ...612L..25N,2017A&A...597A..67A}. However, at the other extreme, the core
composition of ultra-massive (UM) WDs ($M_{\rm WD}/M_{\odot} \gtrsim 1.05$) is uncertain, as they could harbor
CO, oxygen and neon (ONe) \citep{2010A&A...512A..10S,2021A&A...646A..30A}, or even hybrid CO-Ne cores  \citep{2024ApJ...975..259D}.

 From the light that the stars emit, we can learn about the characteristics of their surface layers. However, we can probe the interiors of those stars that show variations in their luminosity due to pulsation modes by means of asteroseismology. This technique compares pulsating stars' observed
frequencies (periods) with those derived by appropriate theoretical models, allowing us to obtain details about their internal structure and evolution. The larger the number of frequencies detected in a given star, the better the asteroseismic analysis becomes.
Until now, five pulsating UM DAV WDs with masses $M_{\rm WD}/M_{\odot} \gtrsim 1.1$ have been reported in the literature. Among them,  BPM~37093 \citep[$T_{\rm eff}= 11\,370$ K, $\log g= 8.84$,][]{2016IAUFM..29B.493N} is the richest pulsating UM WD known so far, with eight pulsation modes detected \citep{2004ApJ...605L.133M} and it has been the target of detailed asteroseismic analysis \citep{2019A&A...632A.119C}.  The remaining objects are GD~518 \citep[$T_{\rm eff}=12\,030$ K, $\log g= 9.084$,][]{2013ApJ...771L...2H},  SDSS J084021.23+522217.4  \citep[$T_{\rm eff}=12\, 160$ K, $\log g= 8.93$,][]{2017MNRAS.468..239C}, WD J212402.03-600100.0 \citep[$T_{\rm eff}=12\, 510$ K, $\log g= 8.98$,][]{2019MNRAS.482.4570G} and WD J004917.14-252556.81 \citep[$T_{\rm eff}=13\,020$ K, $\log g= 9.34$,][]{2023MNRAS.522.2181K}. These objects show, at most, three pulsations modes reported in the literature, preventing comprehensive asteroseismological analyses from being carried out.

We have begun a program aimed at increasing the number of detected UM DAVs and the pulsation modes in already known DAVs with the Gran Telescope Canarias  ({\sl GTC}) and Apache Point Observatory (APO). 
 In this work,  we report the discovery of pulsations in WD~J0135+5722 \citep[$T_{\rm eff}=12\,415$ K, $\log g= 8.90$,][]{2024ApJ...974...12J,2024arXiv241204611K, 2023MNRAS.518.5106J}.  Since we detect an exceptionally high number (19) of pulsation periods, this star constitutes the richest pulsating UM WD known to date. 
The paper is organized as follows. In Sect.~\ref{photometry} we describe the methods applied to obtain the pulsation periods of the target star. In section \ref{evolutionary_models} we derive the stellar mass and the percentage of crystallized core employing different methods. Specifically, we use surface gravity and effective temperature along with evolutionary tracks to derive the spectroscopic mass, and the Gaia parallax and magnitudes to derive the astrometric mass and the photometric mass.
Finally, in Sect.~\ref{conclusions}, we summarize our results.


\section{Candidate selection and photometric observations }
\label{photometry}


To identify UM WD candidates, we utilized the catalog provided by \cite{2023MNRAS.518.5106J}. Our selection criteria included objects with a declination greater than $-10$ degrees, a probability of being DA type greater than 0.5, and masses exceeding  $1 M_{\odot}$ (adjusted from $1.05 M_{\odot}$ to account for potential errors). 
We selected objects that lie within the instability strip defined by \cite{2020NatAs...4..663H}, which is crucial for identifying potential pulsating UM WDs. 
 This process resulted in a total of 19 candidates, including WD~J0135+5722. We supplemented our list with additional ultra-massive DA white dwarfs near the instability strip from \citet[][see their Table 6]{2024ApJ...974...12J}, which includes $M>0.9~M_{\odot}$ white dwarfs within 100 pc and the Pan-STARRS footprint.
It is noteworthy that 14 of the selected UM WDs have been observed with TESS. However, no significant pulsational peaks were detected above the threshold, which is not entirely unexpected given the faintness of these objects.

High-speed photometry was obtained on the night of 2024 Sep 17 with the HiPERCAM imager \citep{2021MNRAS.507..350D} mounted on the 10.4m GTC telescope in La Palma. Data were taken simultaneously in the Super-SDSS $u_{s}/g_{s}/r_{s}/i_{s}/z_{s}$ filters with exposure times of 10.3/2.1/4.1/4.1/4.1 seconds in each band, respectively, with 4x4 binning for a continuous 4h. HiPERCAM uses frame-transfer CCDs so dead time between exposures is negligible.
The HiPERCAM data reduction pipeline was used to debias, flat-field, and fringe subtract the frames before extracting differential photometry relative to a bright comparison star in the field to remove effects due to variations in observing conditions. A variable target aperture size, set to 1.0$\times$FWHM was used to minimize any effects due to the seeing variation and maximize the signal-to-noise ratio of the extracted photometry.

Additionally, high-speed photometry of WD~J0135+5722 on UT 2023 Dec 23 and 2024 Sep 1 was acquired by using the APO 3.5m telescope with the ARCTIC imager and the BG40 filter. Even though the APO observations are relatively short, they confirm multi-periodic oscillations in WD~J0135+5722 over several nights. We obtained back-to-back exposures of 10 s over 92 min on December 23 and 62 min on September 1. To reduce read-out time, we binned the CCD by $3\times3$, resulting in a plate scale of 0.34 arcsec pixel$^{-1}$. This setup has a read-out time of 4.5 s, which results in a cadence of 14.5 s in our light curves.

\subsection{Light curve analysis}

\begin{figure*}
   \includegraphics[width=1.0\textwidth]{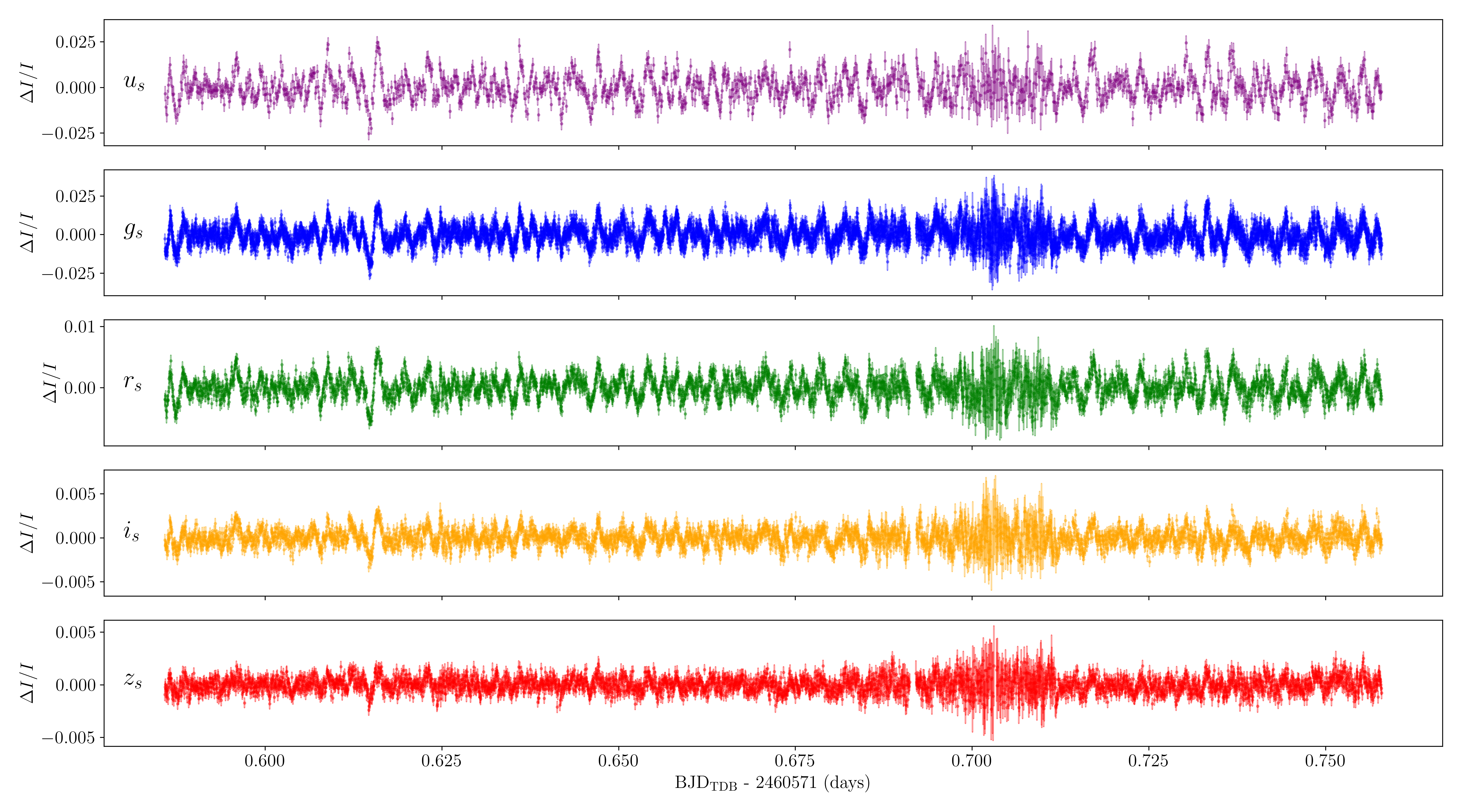} 
 \caption{Light curve obtained with HiPERCAM on September 17 (see
Sec. \ref{photometry}). HiPERCAM filters $u_{s},g_{s},r_{s}
,i_{s},z_{s}$ are shown from top to bottom. The increased scatter at BJD~2460571.7 is due to an increase in the seeing and is unrelated to the target.}
    \label{fig:LC_GTC}
\end{figure*}

\begin{figure}
   \includegraphics[clip,width=1.0\columnwidth]{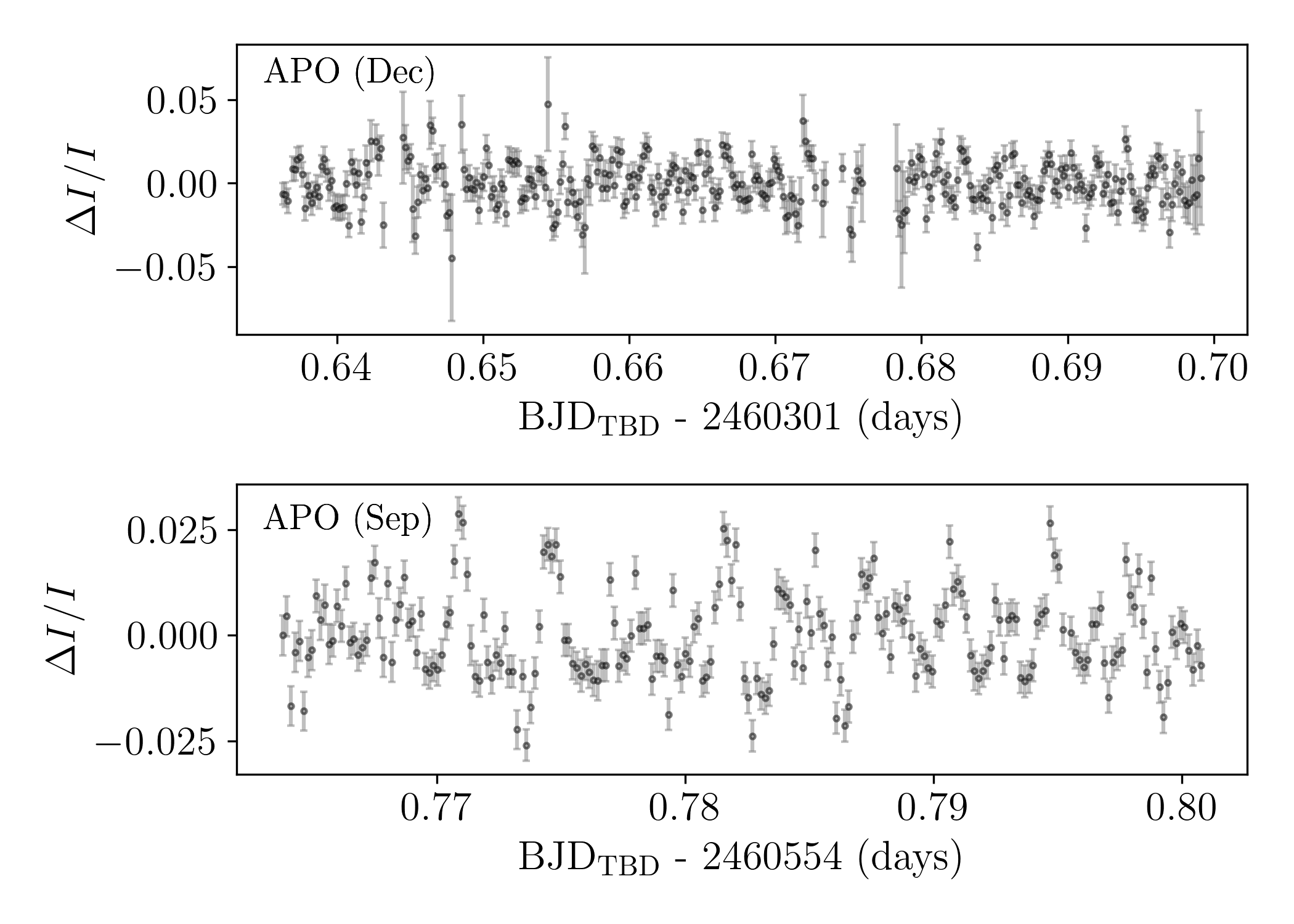} 
 \caption{Light curve obtained with APO on December 23 (upper panel) and September 01 (bottom panel)  (see
Sec. \ref{photometry} for more details).  }
    \label{fig:LC_APO}
\end{figure}

We used a two-step process to analyze the light curves obtained from the HiPERCAM and APO observations, focusing on enhancing the accuracy of the frequency analysis. First, we applied a 5-$\sigma$ clipping method to identify and remove outliers from the data. This technique involves calculating the mean and standard deviation of the flux values and then eliminating data points that deviate significantly (more than 5 standard deviations) from the mean. 
Following the clipping process, we implemented a second-order polynomial detrending to remove long-term systematic variations that could obscure the underlying periodic signals in the light curves. This method involves fitting a polynomial curve to the clipped data and subtracting this fitted curve from the original data, thereby isolating the short-term fluctuations of interest.
The final light curves are shown in Fig. \ref{fig:LC_GTC} for HiPERCAM observations and Fig. \ref{fig:LC_APO} for APO observations. 

\subsection{Frequency analysis}

\begin{figure*}
   \includegraphics[width=1.0\textwidth]{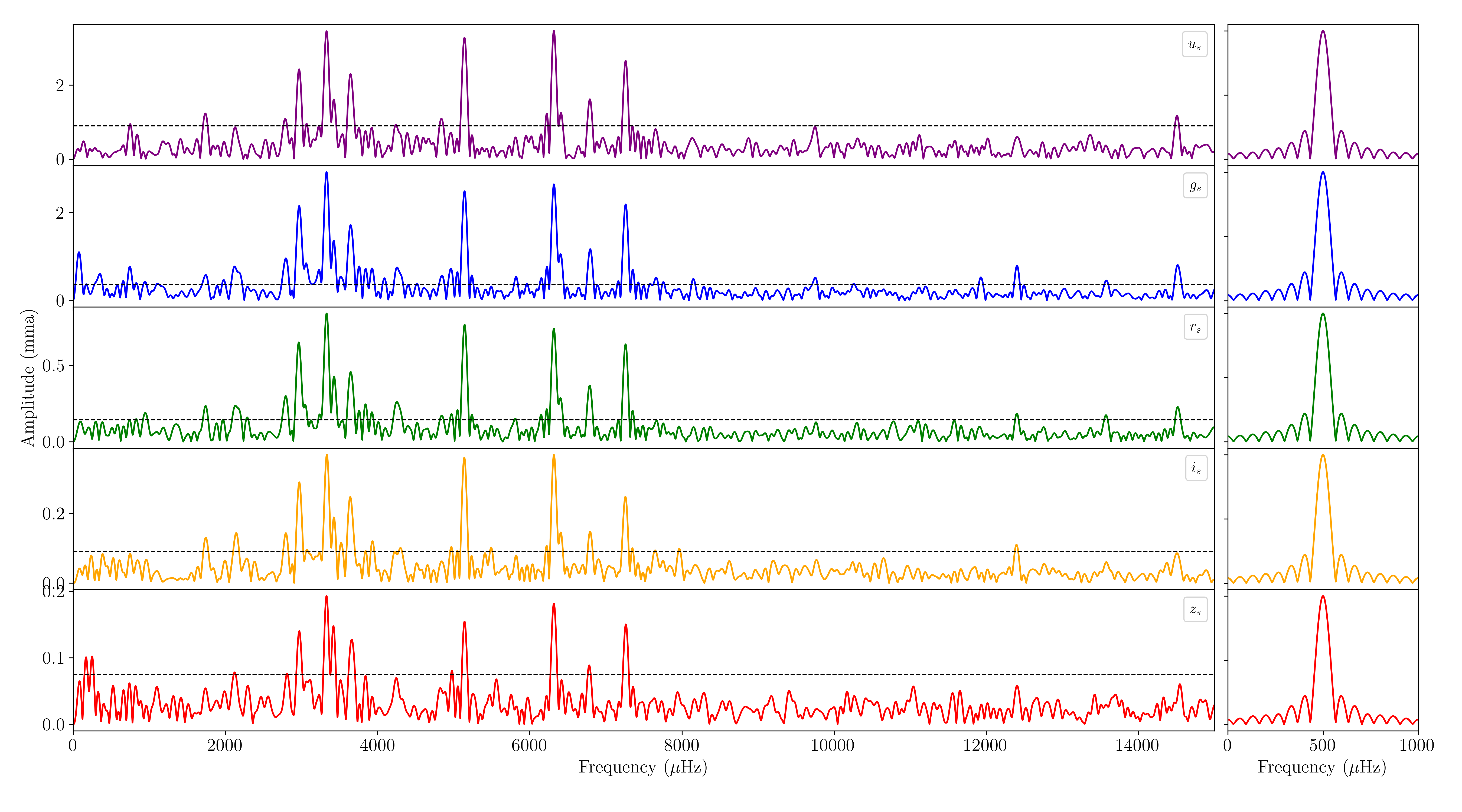} 
 \caption{Fourier transform of the light curves obtained from HiPERCAM, with the same color scheme as in Fig. \ref{fig:LC_GTC}. The horizontal black dashed line indicates the significance level corresponding to S/N=4. The right panels display the spectral window for each light curve centered at 1000 $\mu$Hz for a comparison. }
    \label{fig:FT_GTC}
\end{figure*}

\begin{figure}
   \includegraphics[clip,width=1.0\columnwidth]{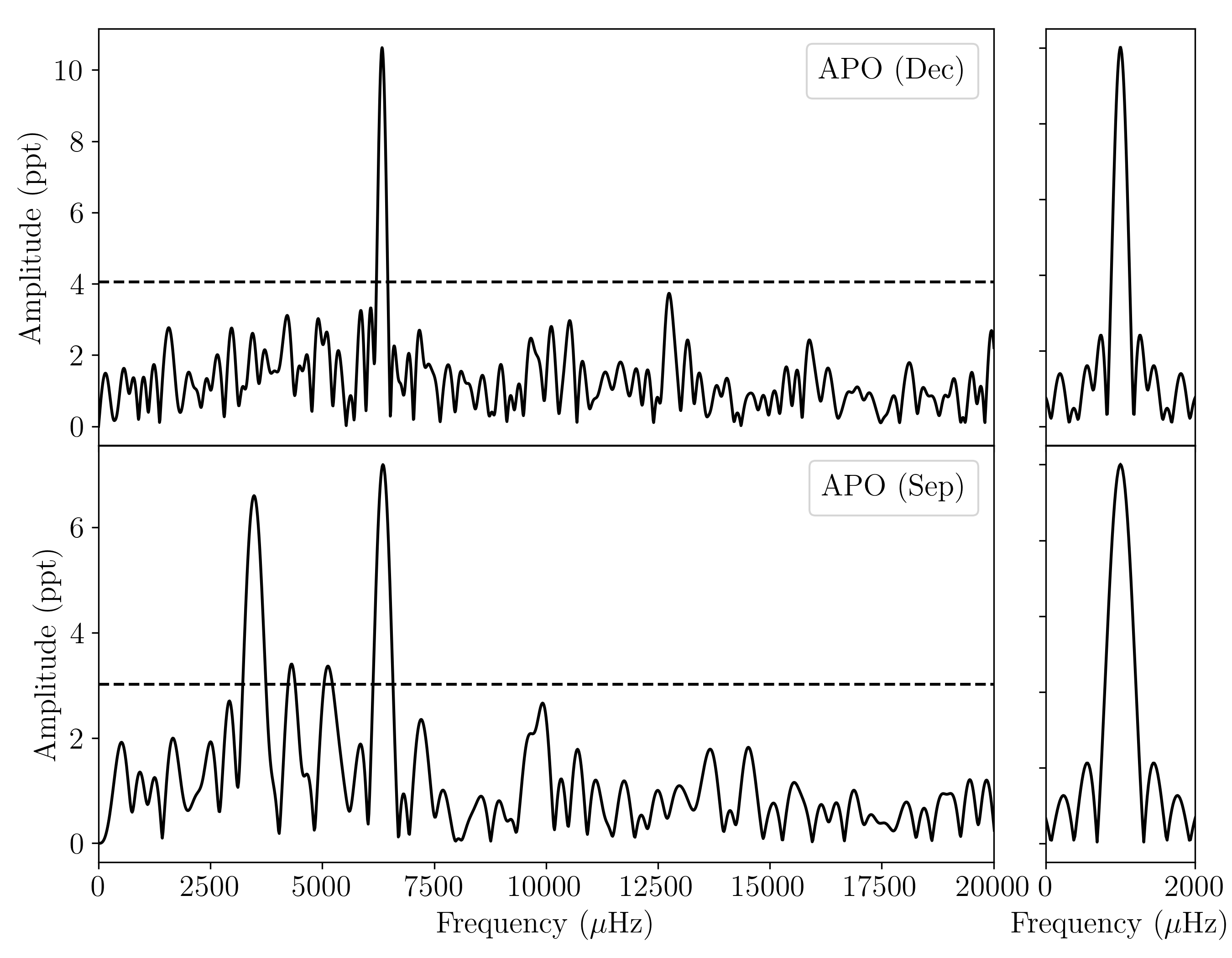} 
 \caption{Fourier transform of the light curves obtained from APO on December 23 (upper panel) and on September 01 (bottom panel). The horizontal black dashed line indicates the significance level corresponding to S/N=4. The right panels display the spectral window for each light curve centered at 1000 $\mu$Hz for a comparison. }
    \label{fig:FT_APO}
\end{figure}

To investigate the periodicities in the light curves of WD~J0135+572, we computed the Fourier Transforms (FT) of each light curve. This analysis allows us to identify the frequencies of the star’s pulsation modes, along with their respective amplitudes, phases and errors. 
The Fourier Transforms for the five bands of HiPERCAM are presented in Fig. \ref{fig:FT_GTC}, while Fig. \ref{fig:FT_APO} illustrates the FT from APO data. 
For each Fourier amplitude spectrum, we computed the median noise level and established the detection threshold, represented by the dashed line in Fig. \ref{fig:FT_GTC} for HiPERCAM and Fig \ref{fig:FT_APO} for APO, using an amplitude ratio of S/N $\ge$ 4, as suggested by \cite{1993A&A...271..482B, 2023ApJS..269...32S}.
For all peaks exceeding this threshold within the frequency resolution of each dataset, we applied a non-linear least squares (NLLS) fit using the software $\tt Period04$\footnote{\url{http://www.period04.net/}} \citep{LB2005}. For some pulsation modes, WD~J0135+572 displayed variations in amplitude, frequency, and/or phase during the observation period, which led to excess residuals in the FT after the process of pre-whitening. We carefully analyzed all frequencies with residuals exceeding the threshold post-prewhitening. When nearby frequencies were within the resolution range, we only fitted and pre-whitened the highest amplitude frequency to avoid confusion with spurious or closely spaced signals. 
We also calculated the spectral window of each light curve shown in the right panels of Fig. \ref{fig:FT_GTC} for HiPERCAM and Fig. \ref{fig:FT_APO} for APO to assess and judge the presence of aliases, which helped us better distinguish between genuine pulsation frequencies and potential artifacts in the data. 
All pre-whitened frequencies for  WD~J0135+572 including both HiPERCAM and APO are given in Table~\ref{table:F_list}, showing frequencies (periods) and amplitudes with their corresponding uncertainties and the S/N ratio.

To conduct a comprehensive analysis of the full FT for each light curve, we first created a combined list from the FT solutions of each light curve. Based on the analysis of all bands ($g_{s}, r_{s}, u_{s}, z_{s}, i_{s}$, APO September, and APO December), several recurring signals are identified, indicating robust frequencies that are consistent across multiple datasets. 
 We present a frequency solution for each subset, including five bands from HiPERCAM and two nights from APO, in Table \ref{table:F_list}, which contains all pre-whitened frequencies for WD~J0135+572 and provides frequencies (periods), amplitudes, associated uncertainties, and the S/N ratio.

To simplify the analysis, we divided the combined FT into three frequency segments, each providing critical insights into the pulsational characteristic of WD~J0135+572. 
In the first segment, we examined frequencies up to 2500 $\mu$Hz, and detected several low-amplitude frequencies.  
The highest peak is located at 77 $\mu$Hz, with a signal-to-noise ratio (S/N) of 7.75. However, this frequency corresponds to 3.6 hours, matching the observing window, and was therefore excluded from the final list. Around 2141 $\mu$Hz, we observe an unresolved peak present in four bands—$g_s, r_s, z_s$, and $i_s$—from observations by the HiPERCAM. This peak is most prominent in the $g_s$ and $r_s$ bands, spanning frequencies between 2128 $\mu$Hz and 2182 $\mu$Hz. From this group, the strongest peak identified is at 2128 $\mu$Hz, with an S/N of 6.5.
Seven peaks were detected in this segment, ranging from 77 $\mu$Hz to 2128 $\mu$Hz, with S/N values between 4.29 and 7.75. 
 
In the second segment, we focused on between $\mu$Hz and 8000 $\mu$Hz. This range contains higher amplitude peaks, where the most prominent signals appear. Most of these peaks are visible across all HiPERCAM observation bands, while only three peaks were detected in the data from the APO, between 3468 $\mu$Hz and 6348 $\mu$Hz. Noteworthy peaks include those around 3846 and 3940 $\mu$Hz, which might be aliases, but they have been retained for future observations to confirm their nature.
An exciting feature is an unresolved, tooth-shaped peak near 4259 $\mu$Hz, where we performed a single extraction. One of the highest peaks in the GTC data appears at 5144 $\mu$Hz, with a nearby peak at 5185 $\mu$Hz in the APO data, suggesting they may correspond to the same mode. 
Additionally, a cluster of peaks near 6330 $\mu$Hz presents a challenge due to slight positional variations between the HiPERCAM and APO datasets. The peak at 6317 $\mu$Hz shows excess residuals, which shift across bands but are likely related. Another peak at 6390 $\mu$Hz, with an S/N of 9.5, was extracted for analysis. Ultimately, the frequencies in this region, including those around 6317, 6335, and 6347 $\mu$Hz, were retained to investigate whether they are part of rotational multiplets.
In total, we identified 15 peaks between 2777 $\mu$Hz and 7259 $\mu$Hz, with S/N values ranging from 5 to 33. 

In the final segment, frequencies beyond 8000 $\mu$Hz exhibit lower S/N values, typically ranging from 4 to 9. The absence of peaks in the z-band or APO September data for this frequency range limits our interpretation, but a few significant peaks were detected between about 12\,400 and 18\,761 $\mu$Hz. The peak at 14\,518 $\mu$Hz has the highest S/N ratio of 9 and is present in almost all bands. The final two peaks, located around 18\,000 and 18\,761 $\mu$Hz, are well-resolved only in the $g_s, r_s, u_s$, and $i_s$ bands. The final list includes 5 significant frequencies from this region. 

We searched for combination frequencies in each band and the APO data using the method described in \cite{2023MNRAS.526.2846U}. Combination frequencies, arising from nonlinear interactions in the surface convection zone, are commonly detected in various pulsating white dwarf classes and provide valuable insights into the thermal response timescale of the convection zone \citep{2005ApJ...633.1142M}. For WD~J0135+572, the FT is inconclusive for identifying all combination frequencies, but we identified two candidates at 12\,406 $\mu$Hz and 14\,518 $\mu$Hz. Given their typically low S/N ratios, confirming these peaks and detecting additional combinations requires a longer baseline. Future observations will be crucial to resolving all combination frequencies beyond 8000 $\mu$Hz.




WD~J0135+5722 presents a compelling challenge for asteroseismic analysis. Despite its rich pulsation spectrum, no clear frequency splitting patterns were detected, complicating the identification of triplets or quintuplets. 

\section{The mass of WD~J0135+5722}
\label{evolutionary_models}

 We estimate the mass of WD~J0135+5722 employing different methods. On the one hand, we calculate the spectroscopic mass using the surface gravity and the effective temperature, together with evolutionary tracks corresponding to ultramassive WD models with different masses and core chemical compositions (ONe and CO). 
We have considered the possibility that WD~J0135+5722 has a core of either ONe or CO because, while it is believed that WD cores in the mass range of WD~J0135+5722 should be composed of ONe \citep{2019A&A...625A..87C,2007A&A...476..893S}, it is not currently ruled out that they could be made of CO \citep[e.g.,][]{2021A&A...646A..30A}. In the case of ONe-core DA WD models, we employ an extension of the model grid of \cite{2019A&A...625A..87C},
which considers the progenitor's complete evolution in the super asymptotic giant branch.  In particular, we consider a grid of model sequences with stellar masses of $M_{\rm WD}/M_{\odot}= 1.10, 1.13, 1.16, 1.19, 1.23, 1.25$, and $1.29$ (De Ger\'onimo et al. 2025, in prep.). In the case of CO-core WD models, we consider
the evolutionary tracks of \cite{2022MNRAS.511.5198C}, which assume the same chemical profiles of C and O in the core for different values of the stellar mass ($M_{\rm WD}/M_{\odot}= 1.10, 1.16, 1.23$, and $1.29$). All our models were generated with the {\tt LPCODE} evolutionary code \citep{2005A&A...435..631A} from the beginning of the WD's cooling track at high luminosities down to the development of the full Debye cooling at very low surface luminosities. 
Additionally, the correct assessment of crystallization and the release of energy due to phase separation are considered. 
The models of ONe-core WDs of \cite{2019A&A...625A..87C} have been successfully employed in previous pulsational studies of H-rich UM WDs \citep[see. e.g.,][]{2019A&A...621A.100D, 2019A&A...632A.119C, 2023MNRAS.522.2181K}.

\begin{figure}
   \includegraphics[clip,width=1.0\columnwidth]{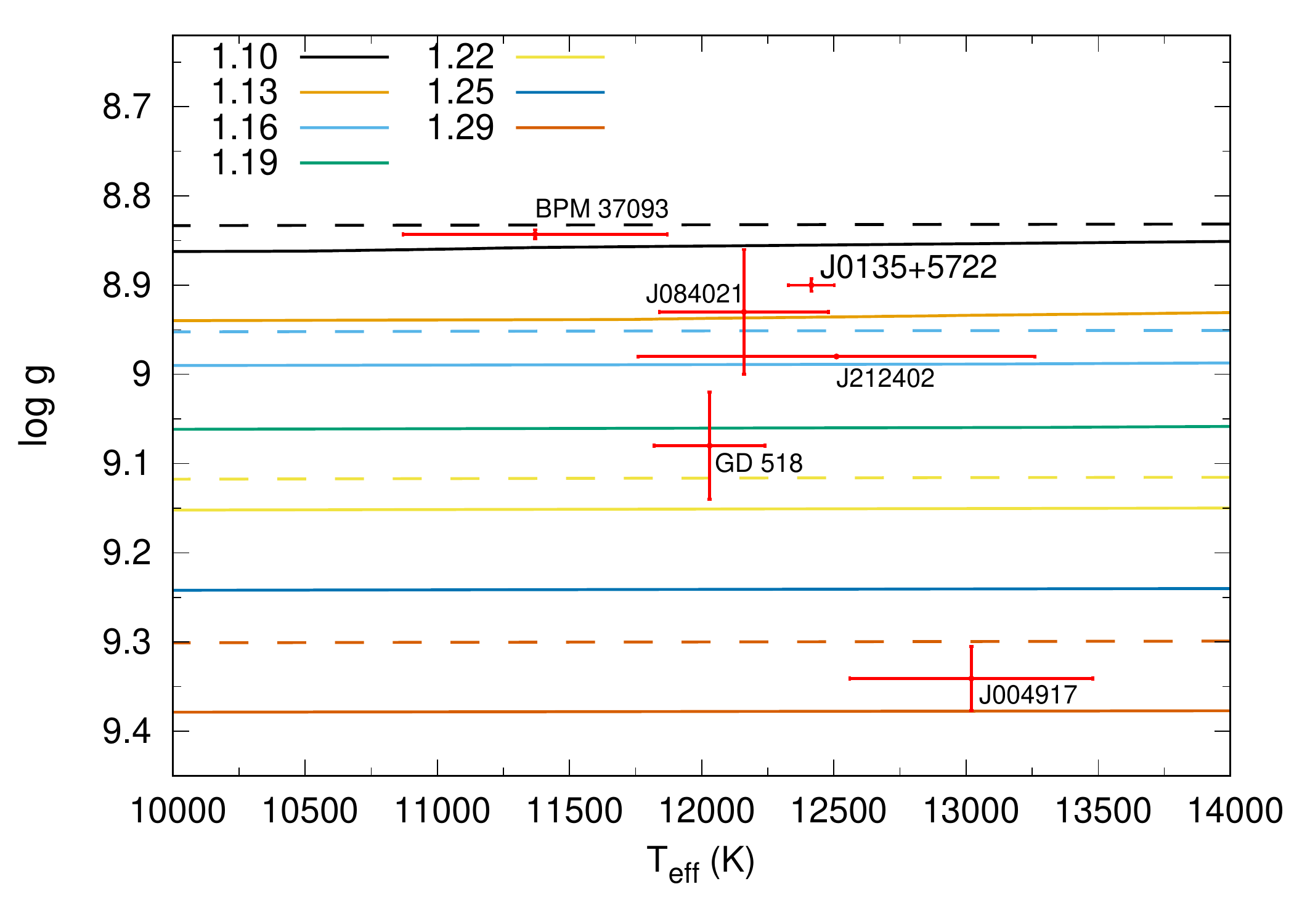} 
 \caption{Location of our UM ONe(CO)-core WD evolutionary tracks, depicted with solid (dashed) lines, together with the determinations for WD~J0135+5722 ($T_{\rm eff}=12\, 415\pm 87$ K and $\log g=8.9\pm 0.007$) in the log $g-T_{\rm eff}$ plane. We also include the spectroscopic determinations for the already known to pulsating UM WDs. The mass of WD~J0135+5722 is $1.118\pm 0.002M_{\odot}$ if its core is made of ONe, 
and $1.135\pm 0.0035 M_{\odot}$ if the star has a CO core.}
    \label{fig:tracks}
\end{figure}

  In Fig. \ref{fig:tracks} we show the evolutionary tracks for our ONe- and CO-core WDs models in the $\log g-T_{\rm eff}$ plane and the spectroscopic determination of the effective temperature and surface gravity of for WD~J0135+5722 from \cite{2024ApJ...974...12J} ($T_{\rm eff}=12\, 415\pm 87$ K and $\log g= 8.90\pm 0.007$).  The error in log $g$ is taken from \cite{2020ApJ...898...84K}. 
 In addition, we include the location of all the previously known pulsating UM ZZ Ceti stars, specifically BPM\,37093 \citep[][]{2016IAUFM..29B.493N}, GD\,518 \citep[][]{2013ApJ...771L...2H},  SDSS J084021.23+522217.4  \citep[][]{2017MNRAS.468..239C}, WD J212402.03$-$600100.0 \citep[][]{2019MNRAS.482.4570G} and WD J004917.14-252556.81 \citep[][]{2023MNRAS.522.2181K}.
 From our evolutionary tracks, we found that, if the star under study harbors a ONe-core, it should have a mass of $1.118\pm 0.002M_{\odot}$.
  On the other hand, if a CO core composition is assumed we found the mass to be $1.135\pm 0.004 M_{\odot}$, slightly lower than the 1.153$M_{\odot}$ predicted by \cite{2024ApJ...974...12J}, who employed CO-core evolutionary sequences from \cite{2020ApJ...901...93B}.
According to our computations, this star would have a crystallized core of about 85.6 (56.0)\% if a ONe(CO)-core composition is assumed.

 The other methods we employed to obtain the stellar mass of WD~J0135+5722 are based on astrometry and photometry. 
 First, we calculated the astrometric mass (which uses Gaia distance and magnitudes together with atmospheric models and evolutionary tracks) and obtained $M_{\rm astr}= 1.137^{+0.004}_{-0.007} M_{\sun}$ (ONe-core WDs) and $M_{\rm astr}= 1.145^{+0.009}_{-0.001} M_{\sun}$ (CO-core WDs)
 following the procedure of \cite{2024A&A...691A.194C}. In this method, distance versus effective temperature curves are calculated for different masses and the stellar mass of the target star is determined by interpolating the value of the Gaia distance of the star. To 
 derive the photometric mass
 \citep[see, e.g.,][]{1997ApJS..108..339B, 2001ApJS..133..413B, 2019ApJ...876...67B,2019MNRAS.482.4570G} we use the Gaia distance together with atmospheric models and mass-radius relations 
 \citep[see][]{2024A&A...691A.194C}, resulting in $M_{\rm phot}= 1.142^{+0.009}_{-0.008} M_{\sun}$ (ONe-core WDs) and $M_{\rm phot}= 1.153^{+0.010}_{-0.009} M_{\sun}$ (CO-core WDs). All stellar mass estimates (spectroscopic, astrometric, and photometric) for WD~J0135+5722 are in excellent agreement. 
 Table \ref{tab:gaia_results} summarizes all solutions derived from various techniques and models. 
 
 Regarding the H content of our ultramassive WD models,  we have adopted the maximum expected to be $\sim 10^{-6}M_{\rm WD}$ for UM WDs. We have found that lower H-mass content could lead to differences in the mass determinations below $\sim 10^{-3}M_{\odot}$. 

 \begin{table}[h!]
    \caption{Main parameters for WD~J0135+5722.}
    \label{tab:gaia_results}
    \centering
    \begin{tabular}{l l}
        \hline
        \multicolumn{2}{c}{Parameters from {\sl Gaia}} \\
        \hline
        Source ID & 412839403319209600 \\
        RA & 01:35:17.57 \\
        Dec & +57:22:49.29 \\
        Parallax & $\pi = 19.665_{-0.057}^{+0.057}$\,mas \\
        Distance & $d = 50.77_{-0.13}^{+0.15}$\,pc \\
        G magnitude & 16.668 \\
        G$_{\text{BP}}$ magnitude & 16.672 \\
        G$_{\text{RP}}$ magnitude & 16.712 \\
        \hline
        \multicolumn{2}{c}{Parameters from spectroscopy} \\
        \hline
        $T_{\rm eff}$ (K) & $12\,415 \pm 87$ \\ 
        $\log g$ (dex) & $8.9 \pm 0.007$ \\ 
        \hline
        \multicolumn{2}{c}{Parameters from evolutionary tracks} \\
        \hline
         $M_{\rm WD}/M_{\odot}$ (ONe-core models) &1.118$\pm $0.002 \\
         $M_{\rm cryst}/M_{\rm WD}$ & 0.85\\
         $M_{\rm WD}/M_{\odot}$ (CO-core models) & 1.135$\pm $0.004\\ 
         $M_{\rm cryst}/M_{\rm WD}$ & 0.56\\
        \hline
        \multicolumn{2}{c}{Parameters from photometry and astrometry} \\
        \hline
        $M_{\rm astr}/M_{\odot}$ (ONe-core models)& 1.137$^{+0.004}_{-0.007}$\\
        $M_{\rm phot}/M_{\odot}$ (ONe-core models) & 1.142$^{+0.009}_{-0.008}$\\
        $M_{\rm astr}/M_{\odot}$ (CO-core models)& 1.145$^{+0.009}_{-0.001}$\\
        $M_{\rm phot}/M_{\odot}$ (CO-core models) & 1.153$^{+0.010}_{-0.009}$\\
        \hline
    \end{tabular}
\end{table}

\section{Conclusions and future prospects}
\label{conclusions}


In this study, we report the discovery of pulsations in WD~J0135+5722, a H-rich UM WD  with 
$T_{\rm eff}=12\,415 \pm 87$ K and $ \log g= 8.90 \pm 0.007$. WD~J0135+5722 displays an exceptionally high number of pulsation modes, surpassing any previously known UM WD. It presents a unique opportunity to investigate the interior of a potential ONe/CO-core WD using asteroseismic techniques.

Using the HiPERCAM and APO instruments, we 
identified 19 distinct pulsation periods, ranging from $\sim$137 s (7259 $\mu$Hz) to 1345 s (743 $\mu$Hz), typical of ZZ Ceti stars. 
This is the sixth pulsating UM WD detected but the richest one in terms of the number of detected frequencies. This discovery is significant for future investigations, as it sheds light on the final stages of high-mass stars and/or merger products, potentially serving as progenitors for supernovae.

WD~J0135+5722 exhibits a complex picture for identifying pulsational modes, with overlapping modes, potential aliases, and deviations from expected patterns, making identifying through both period spacing and rotational splittings more difficult.

We derived a stellar mass of  1.118 or 1.135$M_{\odot}$ if ONe- or CO-core WD evolutionary sequences are adopted, with an 85 or 56\% crystallized core,  respectively. These masses are in excellent agreement with the astrometric and photometric masses derived employing the procedure from \cite{2024A&A...691A.194C}.
UM WDs with masses in the range of masses of WD J0135+572 are 
likely to harbor an ONe-core \citep{2019A&A...625A..87C,2007A&A...476..893S}. However, different evolutionary channels suggest that some UM WD could {\bf have} CO-cores \citep[e.g.,][]{2021A&A...646A..30A}, or even hybrid CO-Ne cores \citep{2024ApJ...975..259D}.
The question of whether this object harbors a core composed primarily of $^{12}$C and $^{16}$O or $^{16}$O and $^{20}$Ne, could be answered by interpreting its pulsation spectrum. For instance, if enough pulsation $g$-modes of consecutive radial order are detected, a study of the forward period spacing could be employed to differentiate the core composition, given that noticeable differences exist in the period spacing 
between ONe-core and CO-core models \citep{2019A&A...621A.100D}. On the other hand, the core chemical composition could be inferred by comparing the observed periods with theoretical periods computed on UM WD models. 
While a detailed set of 
UM WD models with ONe cores are already available \citep[and have been successfully employed in several works, see][]{2019A&A...621A.100D, 2019A&A...632A.119C, 2023MNRAS.522.2181K}, to perform a robust and meaningful seismological analysis of WD~J0135+5722, it is necessary also to consider UM WD models with CO cores. The computation of such a grid of models is currently in progress, and a detailed seismological analysis of WD~J0135+5722 considering  WD models with ONe and CO cores will be the focus of a future paper (De Gerónimo et al 2025, in prep.).

\begin{acknowledgments}
 Based on observations made with the Gran Telescopio Canarias  (Prog. ID: GTC19-24B), installed at the Spanish Observatorio del Roque de los Muchachos of the Instituto de Astrofísica de Canarias, on the island of La Palma.
This work was supported by PIP 112-200801-00940 grant
from CONICET, grant G149 from the University of La Plata, PIP-2971 from CONICET (Argentina) and by
PICT 2020-03316 from Agencia I+D+i (Argentina). 
M. U. gratefully acknowledges funding from the Research Foundation Flanders (FWO) by means of a junior postdoctoral fellowship (grant agreement No. 1247624N). 
This work was partially supported by the MINECO grant  PID2023-148661NB-I00 and by the AGAUR/Generalitat de Catalunya grant SGR-386/2021.
This work is supported in part by the NSF under grant  AST-2205736, the NASA under grants 80NSSC22K0479, 80NSSC24K0380, and 80NSSC24K0436.
The Apache Point Observatory 3.5-meter telescope is owned and operated by the Astrophysical Research Consortium.

This research has made use of the
NASA Astrophysics Data System.
\end{acknowledgments}

%

\vspace{5mm}
\facilities{Gran Telescopio Canarias (GTC), Apache Point Observatory (APO)}




\bibliography{myref}{}
\bibliographystyle{aasjournal}

\appendix
\section{Appendix}

\begin{table*}[ht]
\setlength{\tabcolsep}{8pt}
\renewcommand{\arraystretch}{.9}
\centering
\caption{Frequency solution for WD~J0135+5722.}
\label{table:F_list}
\begin{tabular}{ccccr}
\hline
\noalign{\smallskip}
$\nu$ & $\Pi$ & $A$ & S/N \\
 ($\mu$Hz) & (s) & (ppt) & & \\
\noalign{\smallskip}
\hline
\noalign{\smallskip}

\multicolumn{5}{c}{\textbf{Overall frequency solution}} \\

743.0567	$\pm$  4.183	& 1345.7924	$\pm$  7.5337	 & 0.7122	$\pm$ 0.0801	& 7.91    \\
951.5402	$\pm$  5.5773	& 1050.9277	$\pm$  6.1239	 & 0.2021	$\pm$ 0.0312	& 5.63    \\
1748.4132	$\pm$  5.3341	& 571.9472	$\pm$  1.7396	 & 0.1342	$\pm$ 0.0193	& 5.83    \\
2128.3026	$\pm$  1.1894	& 469.858	$\pm$  0.2624	 & 0.8111	$\pm$ 0.0801	& 9.01    \\
2777.2163	$\pm$  3.7451	& 360.0728	$\pm$  0.4849	 & 0.7967	$\pm$ 0.0801	& 8.85    \\
3331.9871	$\pm$  0.9956	& 300.1212	$\pm$  0.0896	 & 2.9866	$\pm$ 0.0801	& 33.18   \\
3416.0433	$\pm$  3.1816	& 292.7363	$\pm$  0.2724	 & 0.9221	$\pm$ 0.0801	& 10.25   \\
3468.4166	$\pm$  16.0764	& 288.316	$\pm$  1.3302	 & 6.9527	$\pm$ 0.6478	& 9.22    \\
3644.6761	$\pm$  1.5346	& 274.3728	$\pm$  0.1155	 & 1.9304	$\pm$ 0.0801	& 21.45   \\
3846.509	$\pm$  5.6815	& 259.976	$\pm$  0.3834	 & 0.2057	$\pm$ 0.0314	& 5.73    \\
3937.1916	$\pm$  3.8922	& 253.9881	$\pm$  0.2508	 & 0.7529	$\pm$ 0.0801	& 8.37    \\
3940.6543	$\pm$  6.0918	& 253.765	$\pm$  0.3917	 & 0.1175	$\pm$ 0.0193	& 5.11    \\
4259.9674	$\pm$  6.0764	& 234.7436	$\pm$  0.3344	 & 1.2557	$\pm$ 0.0801	& 13.95   \\
4840.4147	$\pm$  6.041	& 206.5939	$\pm$  0.2575	 & 1.1995	$\pm$ 0.1925	& 5.33    \\
5144.2384	$\pm$  1.2187	& 194.3922	$\pm$  0.046	 & 2.4369	$\pm$ 0.0801	& 27.08   \\
5185.0531	$\pm$  27.7865	& 192.8621	$\pm$  1.028	 & 4.0226	$\pm$ 0.6478	& 5.33    \\
6317.6627	$\pm$  1.1259	& 158.2864	$\pm$  0.0282	 & 2.5794	$\pm$ 0.0801	& 28.66   \\
6795.1019	$\pm$  2.4933	& 147.1648	$\pm$  0.054	 & 1.1471	$\pm$ 0.0801	& 12.75   \\
7259.092	$\pm$  1.3482	& 137.7583	$\pm$  0.0256	 & 2.1881	$\pm$ 0.0801	& 24.31   \\

\noalign{\smallskip}
\hline
\noalign{\smallskip}

\multicolumn{5}{c}{\textbf{$u$-band}} \\
1739.051 $\pm$ 5.8416 & 575.0263 $\pm$ 1.9251 & 1.2181 $\pm$ 0.1888 & 5.41 \\
2768.3539 $\pm$ 6.9554 & 361.2255 $\pm$ 0.9053 & 0.9972 $\pm$ 0.1908 & 4.43 \\
2966.8143 $\pm$ 2.9579 & 337.0619 $\pm$ 0.3357 & 2.436 $\pm$ 0.1912 & 10.83 \\
3330.0977 $\pm$ 2.2832 & 300.2915 $\pm$ 0.2057 & 3.5388 $\pm$ 0.1947 & 15.73 \\
3420.9185 $\pm$ 6.6355 & 292.3192 $\pm$ 0.5659 & 1.1771 $\pm$ 0.1951 & 5.23 \\
3642.925 $\pm$ 2.8911 & 274.5047 $\pm$ 0.2177 & 2.4712 $\pm$ 0.1912 & 10.98 \\
4840.4147 $\pm$ 6.041 & 206.5939 $\pm$ 0.2575 & 1.1995 $\pm$ 0.1925 & 5.33 \\
5143.1508 $\pm$ 15.0907 & 194.4333 $\pm$ 0.5688 & 1.6413 $\pm$ 1.8991 & 7.29 \\
6317.0944 $\pm$ 2.0431 & 158.3006 $\pm$ 0.0512 & 3.4252 $\pm$ 0.1903 & 15.22 \\
6791.381 $\pm$ 4.2923 & 147.2455 $\pm$ 0.093 & 1.6115 $\pm$ 0.1906 & 7.16 \\
7258.9402 $\pm$ 2.6556 & 137.7612 $\pm$ 0.0504 & 2.6392 $\pm$ 0.1903 & 11.73 \\
14507.7891	$\pm$  6.0182	& 68.9285 	$\pm$  0.0286	&  1.1632	$\pm$ 0.1894	& 5.17     \\

\noalign{\smallskip}
\hline
\noalign{\smallskip}

\multicolumn{5}{c}{\textbf{$g$-band}} \\
743.0567 $\pm$ 4.183 & 1345.7924 $\pm$ 7.5337 & 0.7122 $\pm$ 0.0801 & 7.91 \\
2128.3026 $\pm$ 1.1894 & 469.8580 $\pm$ 0.2624 & 0.8111 $\pm$ 0.0801 & 9.01 \\
2182.0985 $\pm$ 1.2464 & 458.2745 $\pm$ 0.2616 & 0.6763 $\pm$ 0.0801 & 7.51 \\
2777.2163 $\pm$ 3.7451 & 360.0728 $\pm$ 0.4849 & 0.7967 $\pm$ 0.0801 & 8.85 \\
2972.2266 $\pm$ 1.3874 & 336.4481 $\pm$ 0.157 & 2.1541 $\pm$ 0.0801 & 23.93 \\
3331.9871 $\pm$ 0.9956 & 300.1212 $\pm$ 0.0896 & 2.9866 $\pm$ 0.0801 & 33.18 \\
3416.0433 $\pm$ 3.1816 & 292.7363 $\pm$ 0.2724 & 0.9221 $\pm$ 0.0801 & 10.25 \\
3644.6761 $\pm$ 1.5346 & 274.3728 $\pm$ 0.1155 & 1.9304 $\pm$ 0.0801 & 21.45 \\
3937.1916 $\pm$ 3.8922 & 253.9881 $\pm$ 0.2508 & 0.7529 $\pm$ 0.0801 & 8.37 \\
4259.9674 $\pm$ 6.0764 & 234.7436 $\pm$ 0.3344 & 1.2557 $\pm$ 0.0801 & 13.95 \\
5144.2384 $\pm$ 1.2187 & 194.3922 $\pm$ 0.046 & 2.4369 $\pm$ 0.0801 & 27.08 \\
6317.6627 $\pm$ 1.1259 & 158.2864 $\pm$ 0.0282 & 2.5794 $\pm$ 0.0801 & 28.66 \\
6388.2699	$\pm$  3.0754	&   156.5369	$\pm$   0.0753	    &  0.8525	$\pm$ 0.0801	& 9.47  \\
6795.1019	$\pm$  2.4933	&   147.1648	$\pm$   0.054	    &  1.1471	$\pm$ 0.0801	& 12.75 \\
7259.092	$\pm$  1.3482	&   137.7583	$\pm$   0.0256	    &  2.1881	$\pm$ 0.0801	& 24.31 \\
12406.6927	$\pm$  3.898	&   80.6017	    $\pm$   0.0253	    &  0.7596	$\pm$ 0.0801	& 8.44  \\
14518.184	$\pm$  3.665	&   68.8791	    $\pm$   0.0174	    &  0.809	$\pm$ 0.0801	& 8.99  \\
18001.4722	$\pm$  4.6428	&   55.551	    $\pm$   0.0143	    &  0.6392	$\pm$ 0.0801	& 7.10  \\
18761.3401	$\pm$  4.1854	&   53.3011	    $\pm$   0.0119	    &  0.7093	$\pm$ 0.0801	& 7.88  \\

\noalign{\smallskip}
\hline

\end{tabular}
\end{table*}

\begin{table*}[ht]
\renewcommand\thetable{1}
\renewcommand{\arraystretch}{.9}

\caption{Frequency solution for WD~J0135+5722 (continued).}
\begin{tabular}{ccccr}
\hline
\noalign{\smallskip}
$\nu$ & $\Pi$ & $A$ & S/N \\
 ($\mu$Hz) & (s) & (ppt) & & \\
\noalign{\smallskip}
\hline
\noalign{\smallskip}

\multicolumn{5}{c}{\textbf{$r$-band}} \\

951.5402 $\pm$ 5.5773 & 1050.9277 $\pm$ 6.1239 & 0.2021 $\pm$ 0.0312 & 5.63 \\
1745.0508 $\pm$ 4.7649 & 573.0492 $\pm$ 1.5605 & 0.2430 $\pm$ 0.0310 & 6.77 \\
2135.0815 $\pm$ 4.5913 & 468.3662 $\pm$ 1.0050 & 0.2513 $\pm$ 0.0311 & 7.00 \\
2773.9248 $\pm$ 4.7151 & 360.5000 $\pm$ 0.6117 & 0.2452 $\pm$ 0.0315 & 6.83 \\
2965.5777 $\pm$ 1.8349 & 337.2024 $\pm$ 0.2085 & 0.6488 $\pm$ 0.0316 & 18.07 \\
3332.0720 $\pm$ 1.6993 & 300.1136 $\pm$ 0.1530 & 0.8470 $\pm$ 0.0321 & 23.59 \\
3422.8550 $\pm$ 5.1505 & 292.1538 $\pm$ 0.4390 & 0.2762 $\pm$ 0.0321 & 7.69 \\
3644.7690 $\pm$ 2.2869 & 274.3658 $\pm$ 0.1720 & 0.5217 $\pm$ 0.0315 & 14.53 \\
3846.5090 $\pm$ 5.6815 & 259.9760 $\pm$ 0.3834 & 0.2057 $\pm$ 0.0314 & 5.73 \\
4273.5253 $\pm$ 4.8581 & 233.9988 $\pm$ 0.2657 & 0.2390 $\pm$ 0.0312 & 6.66 \\
5144.3695 $\pm$ 1.5336 & 194.3873 $\pm$ 0.0579 & 0.7558 $\pm$ 0.0311 & 21.05 \\
6317.8238 $\pm$ 2.5577 & 158.2824 $\pm$ 0.0641 & 0.7294 $\pm$ 0.0416 & 20.32 \\
6385.0704 $\pm$ 6.7322 & 156.6153 $\pm$ 0.1650 & 0.2462 $\pm$ 0.0415 & 6.86 \\
6788.5504 $\pm$ 3.1643 & 147.3069 $\pm$ 0.0686 & 0.3584 $\pm$ 0.0313 & 9.98 \\
7259.2770 $\pm$ 1.8073 & 137.7548 $\pm$ 0.0343 & 0.6339 $\pm$ 0.0312 & 17.66 \\
14515.1917 $\pm$ 4.7955 & 68.8933 $\pm$ 0.0228 & 0.2384 $\pm$ 0.0310 & 6.64 \\
17991.8441 $\pm$ 5.7470 & 55.5807 $\pm$ 0.0177 & 0.1996 $\pm$ 0.0310 & 5.56 \\

\noalign{\smallskip}
\hline
\noalign{\smallskip}

\multicolumn{5}{c}{\textbf{$i$-band}} \\

1748.4132 $\pm$ 5.3341 & 571.9472 $\pm$ 1.7396 & 0.1342 $\pm$ 0.0193 & 5.83 \\
2141.8061 $\pm$ 4.7928 & 466.8957 $\pm$ 1.0425 & 0.1493 $\pm$ 0.0193 & 6.49 \\
2784.0118 $\pm$ 6.0437 & 359.1939 $\pm$ 0.7781 & 0.1184 $\pm$ 0.0193 & 5.15 \\
2972.3024 $\pm$ 2.5243 & 336.4395 $\pm$ 0.2855 & 0.2836 $\pm$ 0.0193 & 12.33 \\
3335.4344 $\pm$ 1.9014 & 299.8110 $\pm$ 0.1708 & 0.3765 $\pm$ 0.0193 & 16.37 \\
3422.8550 $\pm$ 5.4009 & 292.1538 $\pm$ 0.4603 & 0.1325 $\pm$ 0.0193 & 5.76 \\
3641.4067 $\pm$ 2.6571 & 274.6191 $\pm$ 0.2002 & 0.2694 $\pm$ 0.0193 & 11.71 \\
3940.6543 $\pm$ 6.0918 & 253.7650 $\pm$ 0.3917 & 0.1175 $\pm$ 0.0193 & 5.11 \\
4300.4239 $\pm$ 7.1133 & 232.5352 $\pm$ 0.3840 & 0.1006 $\pm$ 0.0193 & 4.37 \\
5141.0072 $\pm$ 2.0002 & 194.5144 $\pm$ 0.0756 & 0.3579 $\pm$ 0.0193 & 15.56 \\
6317.8238 $\pm$ 1.9976 & 158.2824 $\pm$ 0.0500 & 0.3583 $\pm$ 0.0193 & 15.58 \\
6391.7951 $\pm$ 6.4168 & 156.4506 $\pm$ 0.1569 & 0.1115 $\pm$ 0.0193 & 4.85 \\
6795.2751 $\pm$ 4.8748 & 147.1611 $\pm$ 0.1055 & 0.1468 $\pm$ 0.0193 & 6.38 \\
7255.9147 $\pm$ 2.8533 & 137.8186 $\pm$ 0.0542 & 0.2509 $\pm$ 0.0193 & 10.91 \\
12396.9219 $\pm$ 6.5691 & 80.6652 $\pm$ 0.0427 & 0.1090 $\pm$ 0.0193 & 4.74 \\
18758.4560 $\pm$ 5.9698 & 53.3093 $\pm$ 0.0170 & 0.1199 $\pm$ 0.0193 & 5.21 \\

\noalign{\smallskip}
\hline
\noalign{\smallskip}

\multicolumn{5}{c}{\textbf{$z$-band}} \\

2972.3024 $\pm$ 4.2728 & 336.4395 $\pm$ 0.4829 & 0.1393 $\pm$ 0.0161 & 7.33 \\
3328.7097 $\pm$ 4.0058 & 300.4167 $\pm$ 0.3611 & 0.1849 $\pm$ 0.0168 & 9.73 \\
3416.1304 $\pm$ 5.7555 & 292.7289 $\pm$ 0.4924 & 0.1237 $\pm$ 0.0168 & 6.51 \\
3654.8560 $\pm$ 4.2284 & 273.6086 $\pm$ 0.3162 & 0.1362 $\pm$ 0.0162 & 7.17 \\
5144.3695 $\pm$ 3.9917 & 194.3873 $\pm$ 0.1507 & 0.1491 $\pm$ 0.0160 & 7.85 \\
6317.8238 $\pm$ 3.2800 & 158.2824 $\pm$ 0.0821 & 0.1803 $\pm$ 0.0161 & 9.49 \\
6781.8257 $\pm$ 6.7705 & 147.4529 $\pm$ 0.1471 & 0.0868 $\pm$ 0.0161 & 4.57 \\
7262.6394 $\pm$ 4.0242 & 137.6910 $\pm$ 0.0763 & 0.1469 $\pm$ 0.0161 & 7.73 \\

\noalign{\smallskip}
\hline
\noalign{\smallskip}

\multicolumn{5}{c}{\textbf{APO (23 December 2023)}} \\

6335.8604 $\pm$ 8.2484 & 157.8318 $\pm$ 0.2052 & 10.5331 $\pm$ 0.8557 & 10.39 \\

\noalign{\smallskip}
\hline
\noalign{\smallskip}

\multicolumn{5}{c}{\textbf{APO (01 September 2024)}} \\

3468.4166 $\pm$ 16.0764 & 288.3160 $\pm$ 1.3302 & 6.9527 $\pm$ 0.6478 & 9.22 \\
5185.0531 $\pm$ 27.7865 & 192.8621 $\pm$ 1.0280 & 4.0226 $\pm$ 0.6478 & 5.33 \\
6347.5099 $\pm$ 14.8682 & 157.5421 $\pm$ 0.3682 & 7.5177 $\pm$ 0.6478 & 9.97 \\

\noalign{\smallskip}
\hline
\end{tabular}
\end{table*}



\end{document}